\documentclass[apj]{emulateapj}
\usepackage{listings}
\usepackage{textcomp}
\usepackage{hyperref}
\hypersetup{
    colorlinks=true,
    linkcolor=blue,
    filecolor=blue,      
    urlcolor=blue,
    citecolor=blue
    }

\usepackage{xcolor}
\usepackage{orcidlink}

\usepackage[utf8]{inputenc}
\usepackage[english]{babel}
\usepackage[T1]{fontenc}

\usepackage{xcolor}
\definecolor{maroon}{cmyk}{0, 0.87, 0.68, 0.32}
\definecolor{halfgray}{gray}{0.55}
\definecolor{ipython_frame}{RGB}{207, 207, 207}
\definecolor{ipython_bg}{RGB}{247, 247, 247}
\definecolor{ipython_red}{RGB}{186, 33, 33}
\definecolor{ipython_green}{RGB}{0, 128, 0}
\definecolor{ipython_cyan}{RGB}{64, 128, 128}
\definecolor{ipython_purple}{RGB}{170, 34, 255}

\usepackage{newunicodechar}
\usepackage[capitalize]{cleveref}
\usepackage{CJKutf8}
\newcommand{\chinesename}{{\begin{CJK}{UTF8}{gbsn}(王加冕)\end{CJK}}}

\DeclareRobustCommand{\okina}{%
  \raisebox{\dimexpr\fontcharht\font`A-\height}{%
    \scalebox{0.8}{`}%
  }%
}
\newunicodechar{ʻ}{\okina}

\usepackage{listings}
\lstdefinelanguage{iPython}{
    morekeywords={access,and,break,class,continue,def,del,elif,else,except,exec,finally,for,from,global,if,import,in,is,lambda,not,or,pass,print,raise,return,try,while},%
    morekeywords=[2]{abs,all,any,basestring,bin,bool,bytearray,callable,chr,classmethod,cmp,compile,complex,delattr,dict,dir,divmod,enumerate,eval,execfile,file,filter,float,format,frozenset,getattr,globals,hasattr,hash,help,hex,id,input,int,isinstance,issubclass,iter,len,list,locals,long,map,max,memoryview,min,next,object,oct,open,ord,pow,property,range,raw_input,reduce,reload,repr,reversed,round,set,setattr,slice,sorted,staticmethod,str,sum,super,tuple,type,unichr,unicode,vars,xrange,zip,apply,buffer,coerce,intern},%
    morekeywords = [3]{>>>} ,
    keywordstyle = [3]\color{black}\bfseries ,
    sensitive=true,%
    morecomment=[l]\#,%
    morestring=[b]',%
    morestring=[b]",%
    morestring=[s]{'''}{'''},% used for documentation text (mulitiline strings)
    morestring=[s]{"""}{"""},% added by Philipp Matthias Hahn
    morestring=[s]{r'}{'},% `raw' strings
    morestring=[s]{r"}{"},%
    morestring=[s]{r'''}{'''},%
    morestring=[s]{r"""}{"""},%
    morestring=[s]{u'}{'},% unicode strings
    morestring=[s]{u"}{"},%
    morestring=[s]{u'''}{'''},%
    morestring=[s]{u"""}{"""},%
    literate=
    {á}{{\'a}}1 {é}{{\'e}}1 {í}{{\'i}}1 {ó}{{\'o}}1 {ú}{{\'u}}1
    {Á}{{\'A}}1 {É}{{\'E}}1 {Í}{{\'I}}1 {Ó}{{\'O}}1 {Ú}{{\'U}}1
    {à}{{\`a}}1 {è}{{\`e}}1 {ì}{{\`i}}1 {ò}{{\`o}}1 {ù}{{\`u}}1
    {À}{{\`A}}1 {È}{{\'E}}1 {Ì}{{\`I}}1 {Ò}{{\`O}}1 {Ù}{{\`U}}1
    {ä}{{\"a}}1 {ë}{{\"e}}1 {ï}{{\"i}}1 {ö}{{\"o}}1 {ü}{{\"u}}1
    {Ä}{{\"A}}1 {Ë}{{\"E}}1 {Ï}{{\"I}}1 {Ö}{{\"O}}1 {Ü}{{\"U}}1
    {â}{{\^a}}1 {ê}{{\^e}}1 {î}{{\^i}}1 {ô}{{\^o}}1 {û}{{\^u}}1
    {Â}{{\^A}}1 {Ê}{{\^E}}1 {Î}{{\^I}}1 {Ô}{{\^O}}1 {Û}{{\^U}}1
    {œ}{{\oe}}1 {Œ}{{\OE}}1 {æ}{{\ae}}1 {Æ}{{\AE}}1 {ß}{{\ss}}1
    {ç}{{\c c}}1 {Ç}{{\c C}}1 {ø}{{\o}}1 {å}{{\r a}}1 {Å}{{\r A}}1
    {€}{{\EUR}}1 {£}{{\pounds}}1
    {^}{{{\color{ipython_purple}\^{}}}}1
    {=}{{{\color{ipython_purple}=}}}1
    {+}{{{\color{ipython_purple}+}}}1
    {*}{{{\color{ipython_purple}$^\ast$}}}1
    {/}{{{\color{ipython_purple}/}}}1
    {+=}{{{+=}}}1
    {-=}{{{-=}}}1
    {*=}{{{$^\ast$=}}}1
    {/=}{{{/=}}}1,
    literate=
    *{-}{{{\color{ipython_purple}-}}}1
     {?}{{{\color{ipython_purple}?}}}1,
    identifierstyle=\color{black}\ttfamily,
    commentstyle=\color{ipython_cyan}\ttfamily,
    stringstyle=\color{ipython_red}\ttfamily,
    keepspaces=true,
    showspaces=false,
    showstringspaces=false,
    rulecolor=\color{ipython_frame},
    frame=single,
    frameround={t}{t}{t}{t},
    framexleftmargin=4mm,
    backgroundcolor=\color{ipython_bg},
    basicstyle=\scriptsize,
    keywordstyle=\color{ipython_green}\ttfamily,
    xleftmargin=1.0em
}

\def\lesssim{\mathrel{\hbox{\rlap{\hbox{\lower4pt\hbox{$\sim$}}}\hbox{$<$}}}}
\def\gtrsim{\mathrel{\hbox{\rlap{\hbox{\lower4pt\hbox{$\sim$}}}\hbox{$>$}}}}

\def\gax{\mathrel{\raise.3ex\hbox{$>$}\mkern-14mu\lower0.6ex\hbox{$\sim$}}}
\def\lax{\mathrel{\raise.3ex\hbox{$<$}\mkern-14mu\lower0.6ex\hbox{$\sim$}}}
\def\gtorder{\mathrel{\raise.3ex\hbox{$>$}\mkern-14mu
             \lower0.6ex\hbox{$\sim$}}}
\def\ltorder{\mathrel{\raise.3ex\hbox{$<$}\mkern-14mu
             \lower0.6ex\hbox{$\sim$}}}

\begin{document}

\title{ASAS-SN Sky Patrol v2.0}

\author{K. Hart\altaffilmark{1,2}}
\author{B.~J. Shappee\altaffilmark{1}\orcidlink{0000-0003-4631-1149}}
\author{D. Hey\altaffilmark{1}}
\author{C.~S. Kochanek\altaffilmark{3,4}\orcidlink{0000-0001-6017-2961}}
\author{K.~Z. Stanek\altaffilmark{3,4}}
\author{L. Lim \altaffilmark{2}}
\author{S. Dodds\altaffilmark{1}}
\author{M. Tucker\altaffilmark{1,3,4}\orcidlink{0000-0002-2471-8442}}
\author{T. Jayasinghe\altaffilmark{3,4,5,6}}
\author{J.~F. Beacom\altaffilmark{3,4,7}}
\author{T. Boright\altaffilmark{8}}

\author{T. Holoien\altaffilmark{9}\orcidlink{0000-0001-9206-3460}}
\author{J. M. Joel Ong \chinesename\altaffilmark{1,6}\orcidlink{0000-0001-7664-648X}}
\author{J.~L. Prieto\altaffilmark{10}}
\author{T.~A. Thompson\altaffilmark{3,4,7}\orcidlink{0000-0003-2377-9574}}
\author{D. Will\altaffilmark{3,4,11}}

\altaffiltext{1}{Institute for Astronomy, University of Hawaiʻi at Mānoa, 2680 Woodlawn Drive, Honolulu, HI 96822} 

\altaffiltext{2}{Department of Information and Computer Science, University of Hawaiʻi at Mānoa, 1680 East-West Road, Honolulu, HI 96822} 

\altaffiltext{3}{Department of Astronomy, The Ohio State University, 140 West 18$^{th}$ Avenue, Columbus, OH 43210} 

\altaffiltext{4}{Center for Cosmology and AstroParticle Physics, The Ohio State University, 191 W. Woodruff Avenue, Columbus, OH 43210}

\altaffiltext{5}{Department of Astronomy,  University of California Berkeley, Berkeley CA 94720}

\altaffiltext{6}{NASA Hubble Fellow}

\altaffiltext{7}{Department of Physics, The Ohio State University, 191 W Woodruff Ave, Columbus, OH 43210}

\altaffiltext{8}{Techbore LLC, 1150 El Cajon, CA, 92021}  

\altaffiltext{9}{Carnegie Observatories, 813 Santa Barbara Street, 
   Pasadena, CA 91101}   
  
\altaffiltext{10}{The Astronomy Nucleus, Universidad Diego Portales, 441 Ejercito Libertador Avenue, Santiago, Chile}

\altaffiltext{11}{Deceased}

\begin{abstract}
  
The All-Sky Automated Survey for Supernovae (ASAS-SN) began observing in late-2011 and has been imaging the entire sky with nightly cadence since late 2017. 
A core goal of ASAS-SN is to release as much useful data as possible to the community. 
Working towards this goal, in 2017 the first ASAS-SN Sky Patrol was established as a tool for the community to obtain light curves from our data with no preselection of targets.  Then, in 2020 we released static $V$-band photometry from 2013--2018 for $\sim 61$ million sources.  
Here we describe the next generation ASAS-SN Sky Patrol, Version 2.0, which represents a major progression of this effort. Sky Patrol 2.0 provides continuously updated light curves for $\sim 111$  million targets derived from numerous external catalogs of stars, galaxies, and solar system objects. We are generally able to serve photometry data within an hour of observation. Moreover, with a novel database architecture, the catalogs and light curves can be queried at unparalleled speed, returning thousands of light curves within seconds.  Light curves can be accessed through a web interface (\url{http://asas-sn.ifa.hawaii.edu/skypatrol/}) or a Python client
(\url{https://asas-sn.ifa.hawaii.edu/documentation}). The Python client can be used to retrieve up to 1 million light curves, generally limited only by bandwidth. This paper gives an updated overview of our survey, introduces the new Sky Patrol, and describes its system architecture. These results provide significant new capabilities to the community for pursuing multi-messenger and time-domain astronomy.
\end{abstract}

\maketitle

\section{Introduction}
The All-Sky Automated Survey for Supernovae (ASAS-SN; \citealp{shappee14, kochanek17}) began observing in 2011 with the mission of identifying bright transients across the whole sky with minimal observational bias. 
With its initial two-camera station at the Haleakala Observatory (Hawaii), ASAS-SN was able to image the visible sky roughly once every 5 days. By late 2013, our team had installed two additional cameras on the Haleakala station, by April 2014 we had installed another two-camera station at the Cerro Tololo International Obervatory (CTIO, Chile), and by mid-2015 we had installed two more cameras at the CTIO station. These additional cameras increased our survey cadence and allowed us to image the entire sky every few nights. In late 2017, we added three additional 4-camera stations, one at McDonald Observatory (Texas), one at the South African Astrophysical Observatory (SAAO, South Africa), and an additional station at CTIO. This geographic redundancy in both the northern and southern hemispheres ensures that we are able to observe the entire sky on a nightly basis.  Our stations are hosted by the Las Cumbres Observatory Global Telescope Network (LCOGT; \citealp{Brown2013}) and we have the ability to download and reduce images within minutes of their exposures.

ASAS-SN units use cooled, back-illuminated  FLI ProLine $2K\times2K$ CCD cameras with 14-cm aperture Nikon telephoto lenses. The units' field-of-view is 4.47 degrees on a side (20 degrees$^2$) with a pixel size of 8.0 arcsec. Ideally, each observation epoch consists of three dithered 90 second exposures, though we are currently averaging 2.7 exposures per epoch due to scheduling and weather events. The two original units used $V$-band filters until late 2018, after which we switched to $g$-band filters. The three newer units have used $g$-band filters from the start. The limiting $V$-band and $g$-band magnitudes are $m \sim 17.5$ and $m\sim18.5$, respectively.

In June 2017, we launched Sky Patrol V1.0; \url{https://asas-sn.osu.edu/}. The goal of Sky Patrol V1.0 was to allow users to request light curves from our image archive. Rather than pre-computing light curves for a set number of targets, Sky Patrol V1.0 provides uncensored light curves for any user-specified coordinates. 
Aperture photometry is performed in real time at the user's requested coordinates using local, nearby stars to calibrate the photometry. 
Though the flexibility of that tool is admirable, its need to compute light curves in real time restricts both the size and frequency of users' queries. 

To partially alleviate these limitations, we pre-computed static $V$-band light curves and released the ASAS-SN Catalog of Variable Stars  (\url{https://asas-sn.osu.edu/variables}; \citealp{Jayasinghe2018, Jayasinghe2019a, Jayasinghe2019b}).  We then expanded these catalogs to include  61.5 million stars across the sky used in the variable star search (\url{https://asas-sn.osu.edu/photometry}; \citealp{Jayasinghe2019c}). We have also created a citizen science program using the Zooniverse\footnote{ Zooniverse: \url{https://www.zooniverse.org/}} platform named Citizen ASAS-SN (\citealt{christy21, christy22}) where, as of Jan 1 2023, $\sim5,300$ volunteers have classified over 1,400,000 ASAS-SN $g$-band light curves through searches for unusual variable stars. 

Finally, in September 2021, we further expanded Sky Patrol V1.0 to not only perform forced-aperture photometry on our reduced images but also to enable users to run aperture photometry on the coadded, image-subtracted data for each epoch with or without the flux of the source on the reference image added.

This paper outlines Sky Patrol V2.0, which seeks to maintain the flexibility of its predecessor (which remains available) while massively improving on it in both speed and scale. V2.0 not only serves pre-computed light curves for a select list of $\sim 111$  million targets, but it also continuously updates the light curves in real time. Like Sky Patrol V1.0, they are uncensored light curves with no deliberate delays in the updates. To do this, we have built a system on top of our automated image processing pipeline that performs photometry and updates our public database as new observations are obtained, calibrated, and reduced. 

In \S2, we describe our imaging pipeline and a number of science products resulting from our survey. In \S3, we discuss our collated target list, their catalogs of origin, and how they can be queried. In \S4, we introduce the Python and web interfaces to our dataset, and provide several usage examples.In \S5, we discuss the photometric properties and precision of this survey tool. In \S6, we discuss the design and metrics of our database. In \S7, we give a short summary and explore the applications of this tool in upcoming research.

\section{Survey Overview}
To date, our high-cadence network of telescopes has produced millions of images. By repeatedly conducting observations in an all-sky fixed grid, we are able to leverage deeply stacked reference images to perform high-precision photometry and optical analysis for a number of different science cases. This section describes our imaging and photometry pipeline, as well as a variety of notable science products resulting from our survey's data.

\subsection{Photometry}
\begin{figure*}
%{\centerline{\psfig{figure=Fields.jpg,width=6.5in}}}
\centering
\includegraphics[width=6.0in]{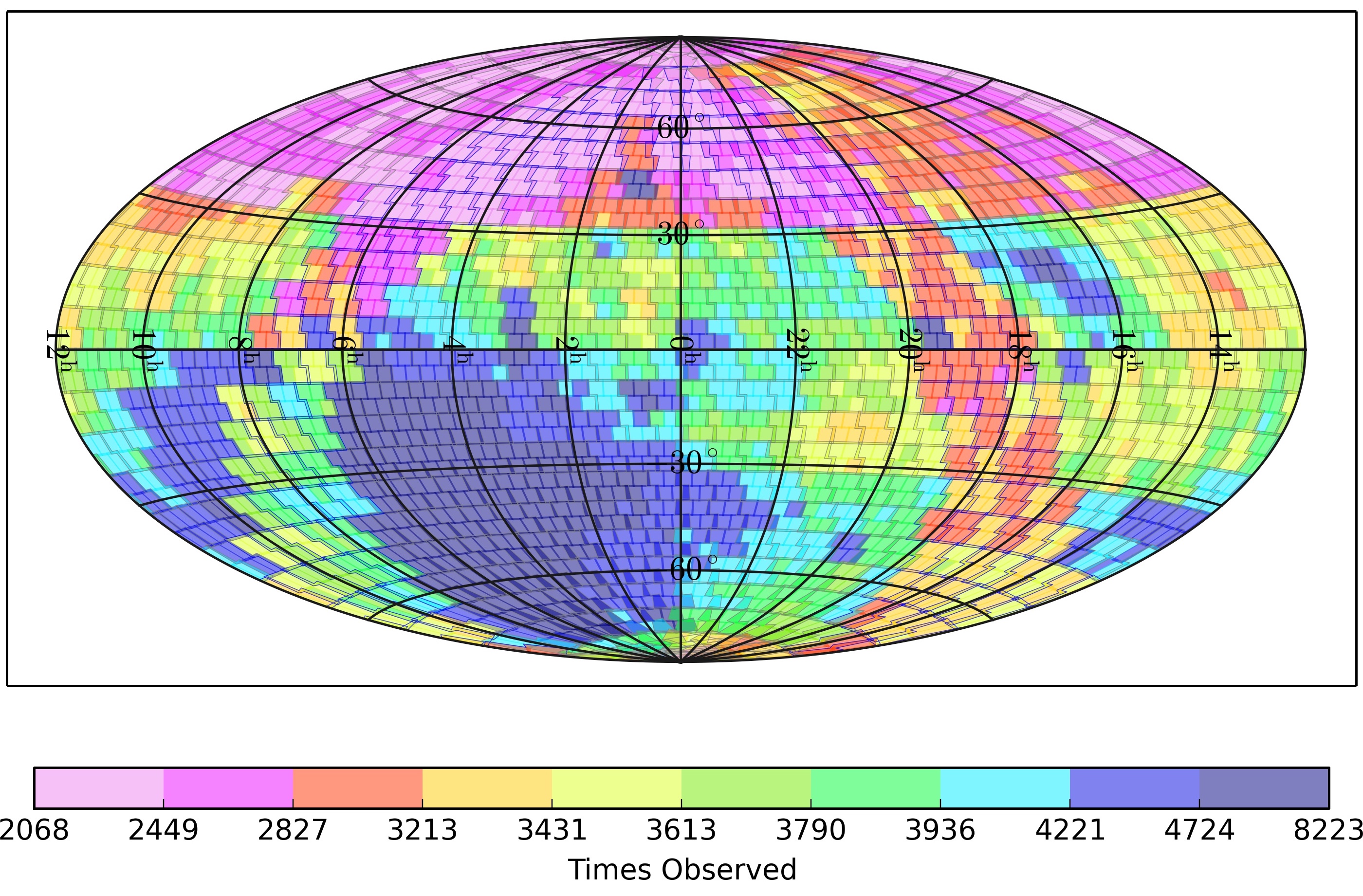}
\caption{Equatorial projection of ASAS-SN fields and number of images taken as of Feb 25, 2023. There are, on average, $2.7$ images for each epoch.
  }
\label{fig:epochs}
\end{figure*}

To achieve full coverage of the sky the ASAS-SN observations are scheduled on 2,824 tessellated fields. Each field matches the 20 degree$^2$ camera field-of-view and has at least a 0.5 degree overlap with adjacent fields. The fields are divided into 706 pointings, where each of the 4 cameras in an ASAS-SN unit observes a specific field. Figure~\ref{fig:epochs} shows the our field map and the number of images taken for each field up to the time of writing.

The data are analysed using a fully-automated pipeline based on the ISIS image-subtraction package \citep{alard98, alard00}. Each photometric epoch (usually) combines three dithered 90-second image exposures with a 4.47$\times$4.47 degree field-of-view of the field that is subtracted from a reference image.  Typically, within an hour of observation our pipeline reduces and co-adds the subtracted images together into observation epochs. We then perform photometry on each of these co-added subtractions and the reference image. 

The reference images are calibrated using isolated stars from the ATLAS Reference Catalog  \citep{tonry18} and by fitting a 2D polynomial to find the zero point as a function of XY position on each reference image. This step reduces any leftover zeropoint issues not removed by flat fielding.\footnote{This calibration scheme is different than the image-subtraction light curves available from Sky Patrol V1.0, which calibrate the reference only using stars near the source.  This means that the light curves provided from Sky Patrol V1.0 and Sky Patrol V2.0 will not be identical.}  
After calibration, we perform photometry on the reference image using \texttt{IRAF apphot} with a fixed aperture of 2 pixels, or approximately 16 arcsec, radius and a background annulus spanning 7--10 pixels in radius.  We perform photometry on all targets from our input catalogs that fall within each field and are more than 0.2 degrees away from the image boundaries.  We then use \texttt{apphot} on the same apertures to perform photometry on each subtracted image, generating a differential light curve to which the reference flux is added.  Finally, the photometry for each target in a given co-add is appended to the corresponding light curves in our database.  

For each epoch in a given light curve, we record the Julian date, camera filter, FWHM, measured flux, flux error, magnitude, magnitude error, and 5 $\sigma{}$ limiting magnitude. For any epoch where the source is not detected at 5 $\sigma{}$, we still give the forced aperture flux but report the limiting magnitude instead of the magnitude and record the magnitude error as 99.99. 

Typically, we publish new photometric measurements within an hour of observation. On the timescale of a day, images are reviewed either manually or by additional quality checks for cirrus, clouds, or other image quality issues.  Images are then determined to be either "good" or "bad."  When this occurs, the corresponding photometry is then flagged as such in the Sky Patrol V2.0 database.  We advise caution when using and interpreting photometry that has not been flagged as "good."

Because we are constantly appending new observations to our light curves and periodically rebuilding reference images---which triggers our pipeline to re-run all photometry on previous co-adds---our light curves are consistently growing in length and improving in quality. 

\subsection{Science Products}
The ASAS-SN survey was initially designed to create a local census of nearby, bright extra galactic transients and has been used to discover and study 
supernovae (e.g., \citealp{shappee16a, holoien16, holoien17, holoien17a, holoien17b, holoien17c, kochanek17b, Bose2019, brown19, holoien19a, shappee19, Vallely2019, Neumann22}; Desai et al. in prep.), 
superluminous supernovae (e.g., \citealp{dong16, godoyrivera17, bose18}), 
rapidly evolving blue transients (e.g., \citealp{Prentice18}), 
tidal disruption events (TDEs; e.g., \citealp{holoien14b, holoien16, holoien16b, holoien18, Holoien2019, Holoien2019b, holoien20, brown16, brown17, hinkle20a, hinkle20b, hinkle21}),
ambiguous nuclear transients (ANTs; e.g., \citealp{neustadt20, hinkle21, holoien22, hinkle22b}),
active galactic nuclei variability (e.g., \citealp{Yuk22}), 
orphan blazar flares  \citep{deJaeger22b},
large outbursts from active galactic nuclei (e.g., \citealp{shappee14, Trakhtenbrot2019, hinkle22, Neustadt22}),
changing-look blazars (e.g., \citealp{mishra21}), and 
even a potential repeating partial TDE (e.g., \citealp{payne20, tucker21, payne22b, payne22}).  
ASAS-SN has also been used to search for optical counterparts for multi-messenger events such as 
high-energy neutrinos observed by Icecube (e.g., \citealp{IceCube17, Garrappa19, IceCube18, franckowiak20, necker22}) and LIGO/VIRGO
gravitational wave events (e.g., \citealp{LIGOCC2020, shappee2019GCNa, shappee2019GCNb, shappee2019GCNc, deJaeger22}).

Moreover, the ASAS-SN data have been used in hundreds of publications to study Galactic and solar system objects.  This includes 
large variable star studies \citep{Jayasinghe2018, Jayasinghe2019a, Jayasinghe2019b, Pawlak2019, Jayasinghe2019c, OGrady20, Jayasinghe21, rowan21, christy22,  christy22b}, 
deriving period--luminosity relationships for $\delta$ Scuti stars \citep{Jayasinghe2019d}, 
fitting detached eclipsing binaries parameters \citep{way22, Rowan22b, rowan22},
studying contact binaries \citep{Jayasinghe2019e},
observing accreting white dwarf systems (e.g., \citealp{Campbell2015, littlefield16}), 
studying young stellar object variability (e.g., \citealp{Holoien2014, herczeg16, Rodriguez2016, Aguilar2017, GullySantiago17, osborn17, Rodriguez2017, bredall20}). 
Furthermore, ASAS-SN was used to examine the long-term variability of Boyajian's Star \citep{Simon2018} and has even recently identified a potential new "Boyajian's Star" analog (the source has been nicknamed Zachy's Star), exhibiting similar rapid dimming events \citep{Way2019}.  ASAS-SN data have also been used to identify 
M-dwarf flares (e.g., \citealp{schmidt14, schmidt16, schmidt19, RodruguezMartinez20, zeldes21}), 
Novae and CVs (e.g., \citealp{kato14, kato14b, kato16, kato17, li17, Aydi19, aydi20, aydi20b, kato20, kawash21, kawash21b, Kawash22}), 
X-ray binaries (e.g., \citealp{tucker18}), 
and R Coronae Borealis stars (e.g., \citealp{shields19}).
ASAS-SN data have also been used to study stars and exoplanets by observing microlensing events (e.g., \citealp{dong19, Wyrzykowski20}), 
determining asteroseismic distances for M giants (e.g., \citealp{auge20}), 
deriving gyrochronologic ages for exoplanet host stars (e.g., \citealp{gilbert20}), 
and vetting exoplanet candidates for background eclipsing binaries (e.g., \citealp{Rodriguez19}).  Finally, ASAS-SN data have even been useful for solar system studies through asteroid shape modeling (e.g., \citealp{Hanus20}), 
discovering 2 new comets in outburst \citep{cometASASSN1, cometASASSN2}, 
and recovering the near-Earth asteroid 2019OK \citep{2019OK}. 

These studies demonstrate the wide utility of the ASAS-SN network and its dataset and why we have created a variety of tools which the community can benefit from.  

\section{Input Catalogs}
Unlike ASAS-SN Sky Patrol V1.0, which allowed users to request light curves at arbitrary points on the sky, Sky Patrol V2.0 by construction only serves pre-computed light curves for objects in our input catalogs. Our input catalogs consist of stellar sources, external catalog sources, and solar system sources. Objects in our stellar source table and catalog source tables have all been cross-matched to a precision of 2 arcseconds and given unique ASAS-SN Sky Patrol Identifiers (\texttt{asas\_sn\_id}\footnote{Not to be confused with ASAS-SN Identifiers given to objects discovered by our survey---e.g., ASASSN-15lh, ASASSN-14li, etc.}).

\begin{table*}
    \centering
\begin{tabular}{|| l | l | r||} 
 \hline
 Source Catalog & Type & \textbf{\textit{n}} sources  \\ [0.5ex] 
 \hline \hline
 ASAS-SN Stellar Source Table & Stellar & 98,602,587   \\
 \hline
 Fermi LAT 10-Year Point Sources & Gamma Ray & 5,788  \\
 \hline
 Chandra Sources v2.0 & X-Ray &  317,224 \\
 \hline
 Swift Master Catalog & Optical/UV/X-Ray/Gamma Ray & 254,936 \\
 \hline
 AllWISE AGN Catalog & Mid-IR/AGN& 1,354,775 \\
 \hline
Million Optical/Radio/X-Ray Associations Catalog (MORX) & Optical/Radio/X-Ray & 3,262,883 \\
 \hline
 Million Quasars Catalog (MILLIQUAS) & QSO & 1,979,676 \\ 
 \hline
 Bright M-Dwarf All Sky Catalog & Stellar & 8,927 \\
 \hline
 AAVSO International Variable Star Index & Stellar & 1,437,528 \\
 \hline
 Galaxy List for the Advanced Detector Era (GLADE) & Galaxy & 3,263,611 \\
 \hline
\end{tabular}
\caption{A breakdown of our input catalogs, with the numbers and types of sources included.}
\label{table:catalogs}
\end{table*}

\subsection{Stellar Sources}
The stellar source table was constructed with the goal of providing an unbiased sample that maximizes the number of targets while maintaining the overall quality of our light curves.  We used the ATLAS Reference Catalog (REFCAT2) to build the target list. Given the pixel size and magnitude sensitivity of our instruments, we chose to include objects with mean $ g < 18.5$ mag and where  r1 $> 20$ arcsec, where r1 is the radius where the cumulative flux exceeds that of the target star. Because REFCAT2 was compiled to include at least 99\% of objects with $m \leq 19$ mag and curated to exclude non-stellar objects, this stellar source table should be considered an exhaustive list of all stars observable by ASAS-SN that are not heavily crowded. 

Because REFCAT2 used Gaia Data Release 2 (Gaia DR2) for its astrometric solutions, we were able to directly match Gaia source IDs from the given RA and DEC coordinates. We then used the best-neighbour tables in the Gaia Archive to cross-match with AllWISE, SDSS, and 2MASS. Finally, we used the same Gaia source IDs to get TESS Input Catalog Version 8.0 (TIC) IDs. This means that users can query our stellar source table using either the original columns provided by REFCAT2 or by providing external catalog identifiers for any of these other cross-matched surveys.

We note that some of our external source catalog tables contain stellar objects that do not occur in the stellar table. However, given the completeness of REFCAT2, these objects typically fall outside our sensitivity or suffer from significant crowding.

\subsection{Additional Catalog Sources}
In addition to our stellar sources, Sky Patrol V2.0 provides light curves for a number of specialized  catalogs from NASA's  High Energy Astrophysics Science Archive Research Center (HEASARC), as shown in Table \ref{table:catalogs}.  With the explicit goal of aiding multi-messenger astronomy we have included the entire source catalogs from Fermi LAT (\citealp{Abdollahi_2020}), Chandra (\citealp{Evans_2010}), and Swift (\citealp{2018yCat....102039N}). For researchers interested in specific object and variable types we have included the AllWISE AGN catalog (\citealp{2015ApJS..221...12S}), the Million Quasar Catalog (Milliquas; \citealp{2021yCat.7290....0F}), the Brown Dwarf Catalog (\citealp{2011AJ....142..138L}), the AAVSO International Variable Star Index (\citealp{2006SASS...25...47W}) and the  Galaxy List for the Advanced Detector Era (GLADE; \citealp{Dalya2018}). 

Whereas our stellar sources were selected from REFCAT2 to ensure detections given our magnitude limits, these catalogs were not pruned based on flux. Photometry is performed at all target coordinates in these catalogs without bias. As with the stellar sources, null detections will be timestamped and reported in our light curves with $\sigma \sim 99.99$. Our interface allows the user to query sources using all of the original columns of the input catalogs and, we have maintained the original naming conventions of the catalogs' columns with few exceptions. 

Because many sources appear in more than one of these catalogs and we have already done the cross-matching, users can perform JOIN operations on these tables using \texttt{asas\_sn\_id}s. In \S5, we provide several examples of such queries in ADQL.

Finally, for a few of the catalogs, ASAS-SN spatial resolution is better than the accuracy of the catalog's coordinates for some of its objects.  In these cases, we make no effort to identify the corresponding optical counterpart and simply perform forced aperture photometry at the catalog's reported coordinates with our normal aperture size.  Thus, users must be cautious when using the ASAS-SN Sky Patrol V2.0 photometry for poorly localized sources.  In these cases we recommend first searching for the corresponding optical counterpart's photometry in Sky Patrol V2.0, and if not present, to use the slower Sky Patrol V1.0 to compute forced aperture photometry at the exact coordinates desired. 

\subsection{Solar System Sources}
To provide photometry for asteroids and comets, we rely on updates from the Minor Planet Center Orbit Database (MPCORB Database) every 10 days.  For each object in the most recent MPCORB Database relative to our observation, we calculate the position and run aperture photometry on all objects within that image. Unlike our other sources, we do not add flux from the reference images. Photometry on asteroids is run with the same aperture as our extra-solar sources, while comets are run in a different manner. Comet photometry is computed using 2--8 pixel apertures and a 15--20 pixel annulus. This means that each comet's light curve will have 7 different magnitude values for each observation epoch. The goal of this regime is to provide  researchers a detailed picture of the growth and decay of gas and dust tails as these comets traverse our solar system.

\section{Interface}
We have designed a simple, yet powerful interface for users to query objects from our input catalog tables and to seamlessly retrieve their corresponding light curves. This interface can be accessed through our Python client or our web portal. While these have similar functionality, queries on the web portal will have stricter limits on the number of sources returned in a single query.  The web portal limits users to 10,000 sources per query, whereas the Python client allows up to 1 million sources per query.  The latest version of the Python tool and documentation can be found at \url{https://asas-sn.ifa.hawaii.edu/documentation}. The web portal can be accessed at \url{https://asas-sn.ifa.hawaii.edu/skypatrol}. 

ASAS-SN Sky Patrol light curves can be queried in a number of ways: cone searches, catalog IDs, random samples, and ADQL queries. Further statistical and plotting tools are also provided by the Python client. We detail the available methods of querying light curves below.

\subsection{Cone Search}
The cone search is the most basic operation that we allow. Users can provide RA, DEC, and a search radius, and obtain catalog entries and light curves for all sources within that cone. Our system is unique in that there is no limit on the radius of a cone search and that the speed of the query will not be affected by the size of the cone. Users can also filter the desired sources in their cone search by catalog. Moreover, our ADQL function set includes utilities for running cone searches in conjunction with more complex queries. 

\begin{lstlisting}[language=iPython]
>>> from pyasassn.client import SkyPatrolClient
>>> client = SkyPatrolClient()
>>> client.cone_search(ra_deg=277.0, 
                       dec_deg=-12.1, 
                       radius=5.0, 
                       units='arcmin', 
                       catalog='aavsovsx')
                       
asas_sn_id    ra_deg     dec_deg    name
292059085873  276.84665 -12.01138   GDS_J1827232
326417718321  277.36058 -12.37650   ASASSN-V J18
395137087890  280.42779 -10.01075   NSVS 1675161
...            ...        ...       ...                   
[6594 rows x 4 columns]
\end{lstlisting}

\subsection{Catalog IDs}
Searching by catalog ID is the simplest way to access our data. To do this, the user can provide target IDs for any catalog that we have cross-matched against and our interface will return light curves for all the sources in our data that match that list. Our stellar source table can be queried with TESS, GAIA DR2, AllWISE, SDSS, REFCAT2, and 2MASS identifiers. Sources from the other catalogs can be queried directly by their survey identifiers (e.g., Swift, Chandra, or Fermi LAT).

\begin{lstlisting}[language=iPython]
>>> my_tic_ids = [6658326, 46783395, 1021890]
>>> client.query_list(my_tic_ids, 
                      catalog='stellar_main', 
                      id_col='tic_id') 
                       
asas_sn_id    ra_deg      dec_deg      tic_id
309238124040  329.260377  -8.035864   1021890
335007699083   97.045759  18.214838  46783395
335007693701   81.164422  18.222147   6658326
\end{lstlisting}

This utility is meant for users that already have a source list from a specific survey and that hope to use our light curves to supplement their research. This is also the fastest way to query our database.

\newpage
\subsection{Random Samples}
This utility has been included for data scientists looking to train or test models on large unbiased samples of light curves. The random sample function allows users to pull arbitrary numbers of light curves along with their corresponding catalog data. The following example shows a random sample from the MORX catalog.  This utility returns a new random sample each time. If users want to perform multiple experiments on the same random sample, they will need to save the returned \texttt{asas\_sn\_id}s.  

\begin{lstlisting}[language=iPython]
>>> client.random_sample(10000, catalog="morx")
                       
asas_sn_id    ra_deg      dec_deg   name
661431605389  348.28629  -89.26930  IRAS 22416-
661431261661  209.09643   40.16277  SDSS J13562
77310593719   307.30222   40.16305  MORX J20291
17181160891   257.67916  -87.50250  TYCHO 9530-
...            ...        ...       ...                   
[10000 rows x 4 columns]
\end{lstlisting}

\subsection{ADQL Queries}

\begin{figure*}[t!]
\includegraphics[width=7in]{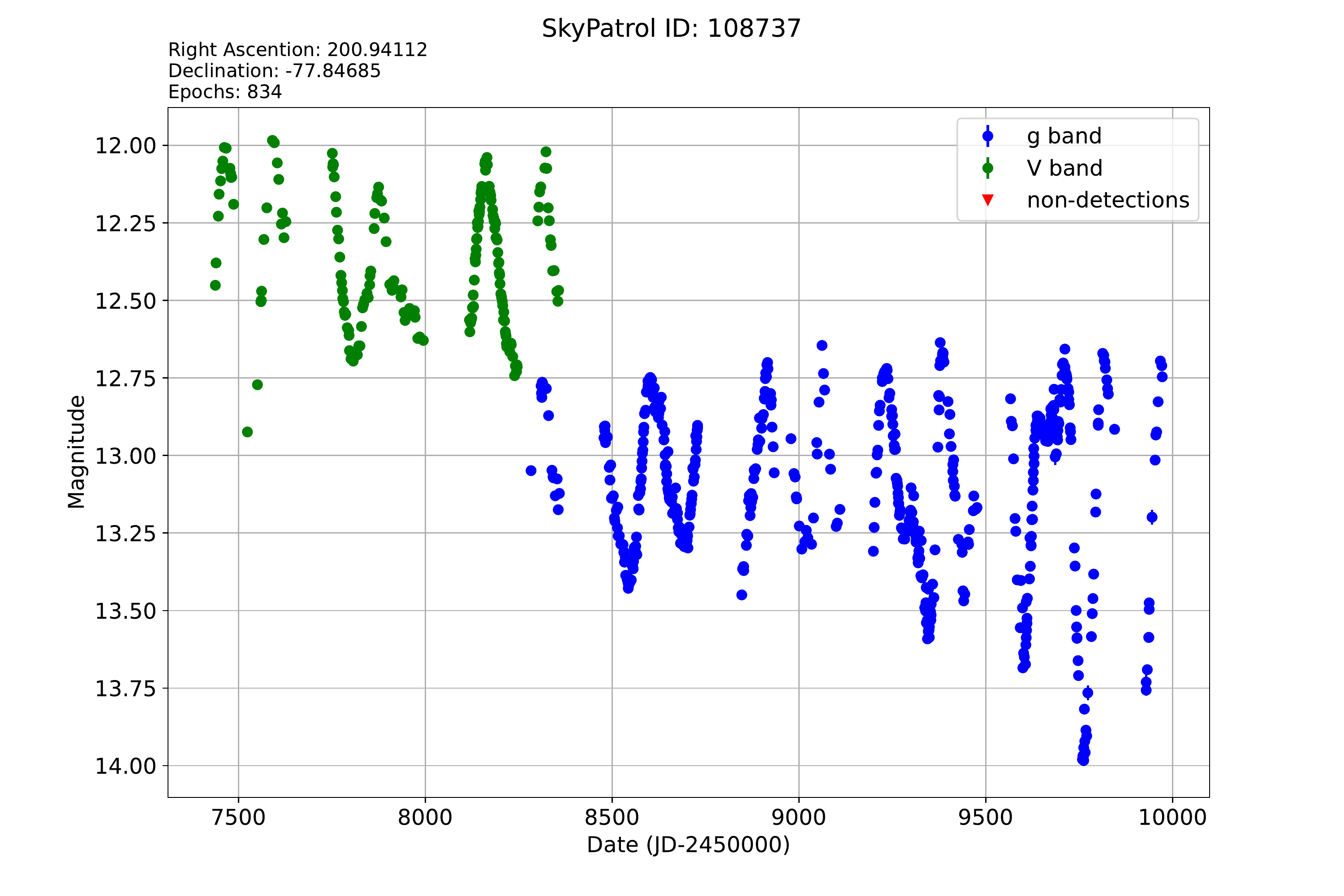}
\caption{Sample light curve generated by the Sky Patrol V2.0 Python client for the long-period variable star, BP Cha. The plot includes basic metadata as well as photometry for both g- and V-band filters.}
\label{fig:bp_cha}
\end{figure*}

\begin{figure*}[t]
\centering
\includegraphics[scale=1.2]{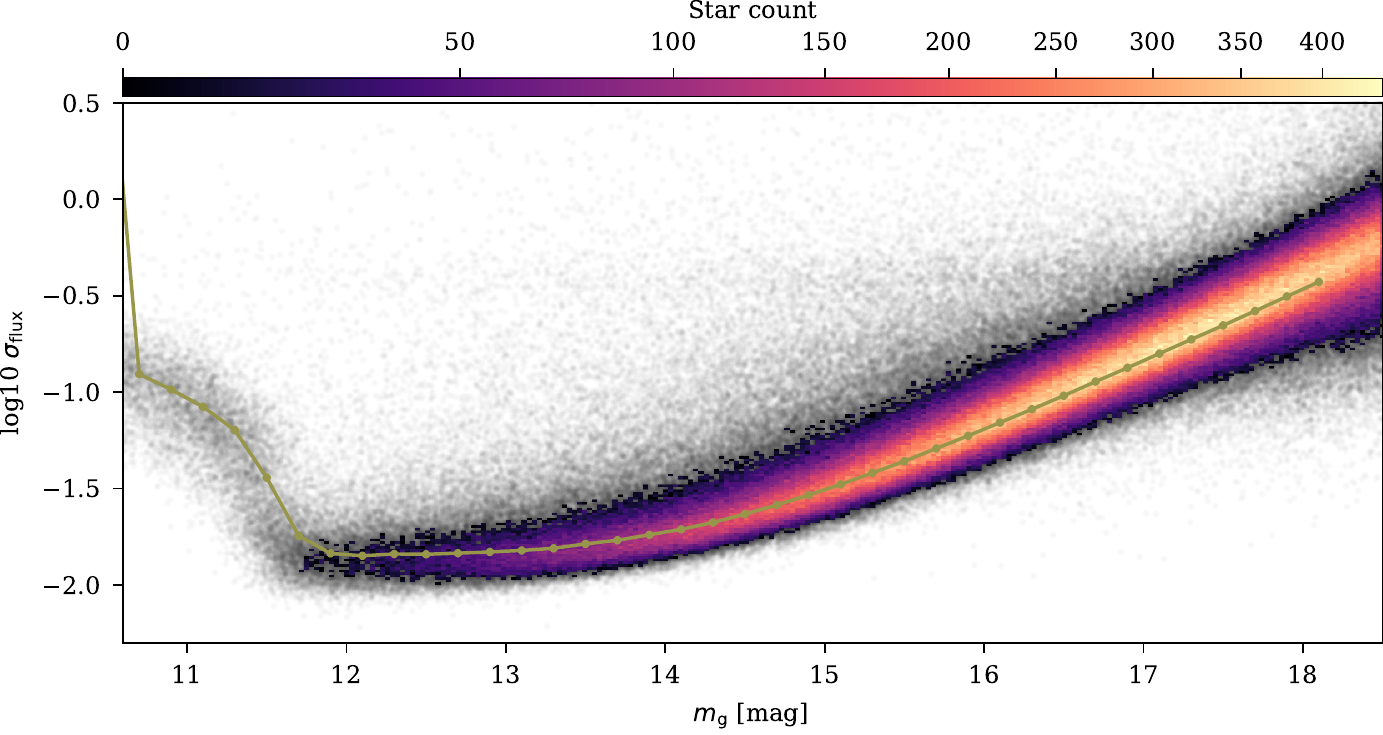}
\caption{Photometric precision of the $g$-band photometry as a function of Pan-STARRS $g$ magnitude. The dotted line represents the 50th percentile of the standard deviation of the flux in each bin. Note that saturation begins to set in around $g \sim 12$th mag.}
\label{fig:sat_range}
\end{figure*}

\begin{figure}
\includegraphics[width=\linewidth]{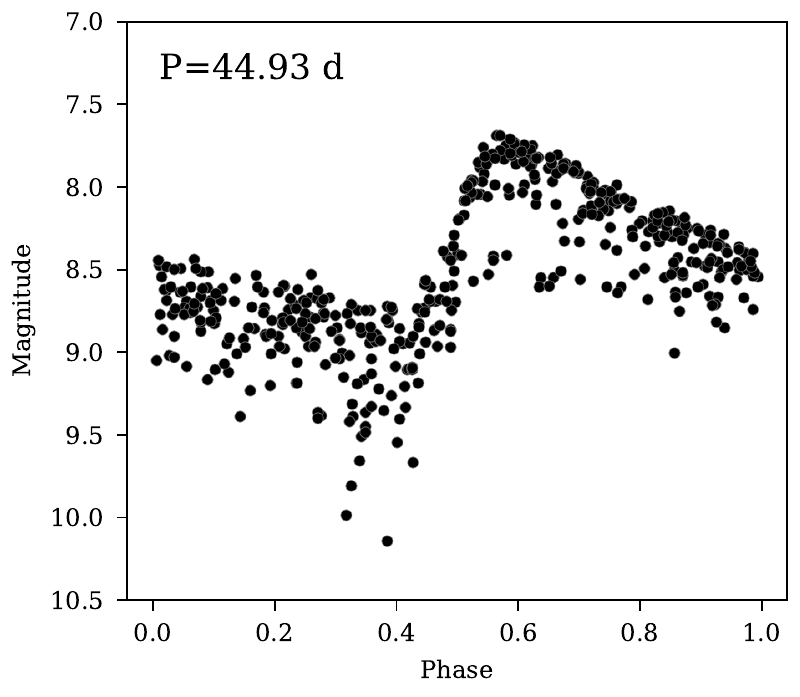}
\caption{Phased $g$ band light curve of the bright classical Cepheid SV Vul. Despite the extreme saturation, the photometry is still sufficiently clean for the periodicity to be measured.}
\label{fig:saturated_ceph}
\end{figure}

While many other survey utilities, such as the Gaia DR2 Archive and VizieR, run interfaces compliant with the International Virtual Observatory Alliance's (IVOA) Astronomical Data Query Language  (ADQL) specification, we have chosen to use a custom grammar that both adds functionality and simplifies geometric queries\footnote{The full query syntax can be found at \href{https://asas-sn.ifa.hawaii.edu/documentation/additional.html\#syntax}{https://asas-sn.ifa.hawaii.edu/documentation/additional.html\#syntax}}.  

In addition to the query functionality of traditional ADQL, we have included support for Common Table Expressions, \textbf{WINDOW} functions, correlated subqueries, and \textbf{UNIONS}. We have removed the common geometry functions such as \textbf{BOX}, \textbf{CIRCLE}, \textbf{AREA}, \textbf{POINT}, and \textbf{CONTAINS}. Instead of building predicates out of these functions, users can perform geometric searches either using the \textbf{DISTANCE} function or by specifying range conditions on RA and DEC. Below is an example cone search\footnote{Additional query examples can be found at \href{https://asas-sn.ifa.hawaii.edu/documentation/additional.html\#example-queries}{https://asas-sn.ifa.hawaii.edu/documentation/additional.html\#example-queries}}.

\begin{lstlisting}[language=iPython]
>>> query = """
SELECT 
    asas_sn_id, ra_deg, dec_deg
FROM stellar_main 
WHERE DISTANCE(ra_deg, dec_deg, 270, -88) <= ARCMIN(30)
"""
\end{lstlisting}

\begin{lstlisting}[language=iPython]
>>> client.adql_query(query)
                       
asas_sn_id      ra_deg    dec_deg
1167369      262.654795 -88.335208
1167713      275.492788 -88.337208
17181154501  271.738674 -88.263265
...            ...        ...                 
[2321 rows x 3 columns]
\end{lstlisting}

Because our data is stored in decimal degrees, we have provided functions to enter coordinates in HMS and DMS formats as well as functions to give distance constants in minutes and seconds of arc.~%\footnote{These functions can be found at \href{https://asas-sn.ifa.hawaii.edu/documentation/additional.html#functions}{https://asas-sn.ifa.hawaii.edu/documentation/additional.html\#functions}}. 
Furthermore, because our lookup tables are running on top of Apache Spark, users can also leverage all of the Spark SQL functions\footnote{A comprehensive list of these functions can be found at \url{https://spark.apache.org/docs/latest/api/sql/index.html}}. 

The ADQL interface can be queried at will and used as a tool for exploring our source catalogs without any requirement that the user downloads the corresponding light curves. The speed of our execution engine allows scientists to query data on multiple sources across our input catalogs. However, users interested in the light curves for these catalog sources must include \texttt{asas\_sn\_id} as a column name in order to download the light curves.

\subsection{Light Curve Utilities}
The Python client provides additional functionality for downloading and processing individual or collections of light curves.

\subsubsection{Collections}
Once the user has found a set of targets through any of the 4 catalog query functions, they can now download their light curves. The cone search, catalog id, random sample, and ADQL query functions all have a boolean download parameter. If set, then the query function will return a \textbf{LightCurveCollection} object. This object provides the user with aggregate statistics on the collected light curves.

\begin{lstlisting}[language=iPython]
# Wide angle cone search of targets near the pole
>>> lcs = client.cone_search('18:54:11.5', 
                             '-88:02:55.22', 
                             radius=2.0,
                             units='deg',
                             download=True, 
                             threads=8)
>>> lcs.stats()
                    
asas_sn_id    mean_mag    std_mag   epochs           
443093        11.133392  0.289588     416
443275        16.360747  0.129060     394
443276        15.826793  0.097284     412
...            ...        ...       ...
[30693 rows x 4 columns]
\end{lstlisting}

\subsection{Individual Light Curves}
Once the user has downloaded a collection, they can view the individual light curves, as well as their meta data, periodograms, and plotted light curves. Moreover, because light curve data is held in memory as a \textsc{pandas} DataFrame, they can easily be saved to disk in csv format. Individual curves are retrieved from the collection using their \texttt{asas\_sn\_id}.

\begin{lstlisting}[language=iPython]
>>> lightcurve = lcs[108737]
>>> lightcurve.data

     jd   flux flux_err mag mag_err limit  fwhm
2458680  18.53  0.10  13.23  0.02  17.09  1.45
2459351  14.42  0.12  13.50  0.01  16.87  1.65  
2458706  18.52  0.11  13.23  0.01  16.97  1.43
2459308  18.66  0.09  13.22  0.02  17.23  1.47
2458925  26.60  0.12  12.83  0.00  16.95  1.42   
[834 rows x 12 columns]
\end{lstlisting}

As well as providing the photometric data for each of our light curves. We also provide some basic functions, such as plotting, lomb-scargle, and period finding. Fig.~\ref{fig:bp_cha} shows a sample plot for a long period variable star from the AAVSO catalog.

\begin{lstlisting}[language=iPython]
# Simple Plot
>>> lightcurve.plot(phot_filter='all')

# An Astropy LombScargle object is also returned
>>> f, p, ls = lightcurve.lomb_scargle()
>>> lightcurve.find_period(f)
152.32918886138538
\end{lstlisting}

\section{Photometric properties and examples}

 \begin{figure}
\includegraphics[width=\linewidth]{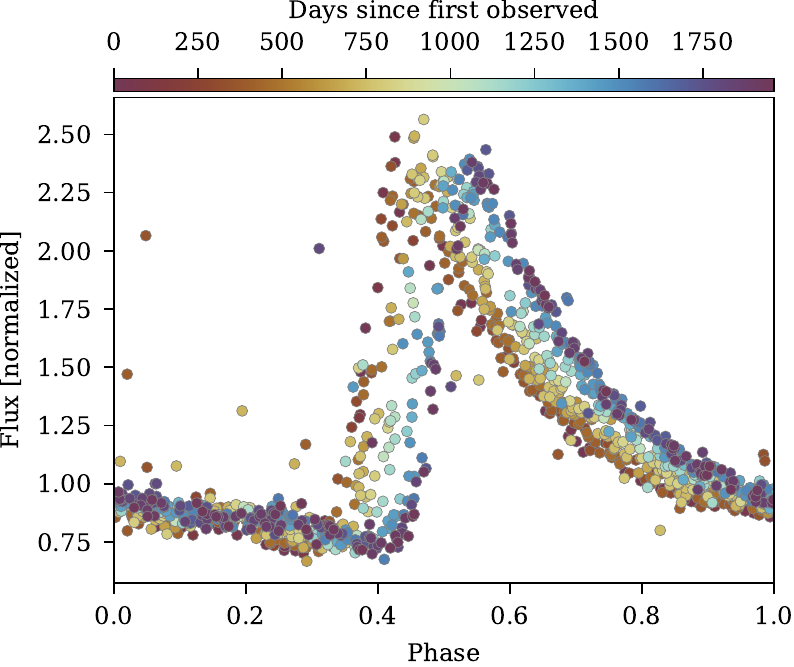}
\caption{Long period Blazhko modulation in the RR Lyrae variable CD Scl observed by ASAS-SN in $g$-band. The long baseline of g-band data allows for multiple cycles to be observed.}
\label{fig:RRLyr}
\end{figure}

\begin{figure}
\includegraphics[width=\linewidth]{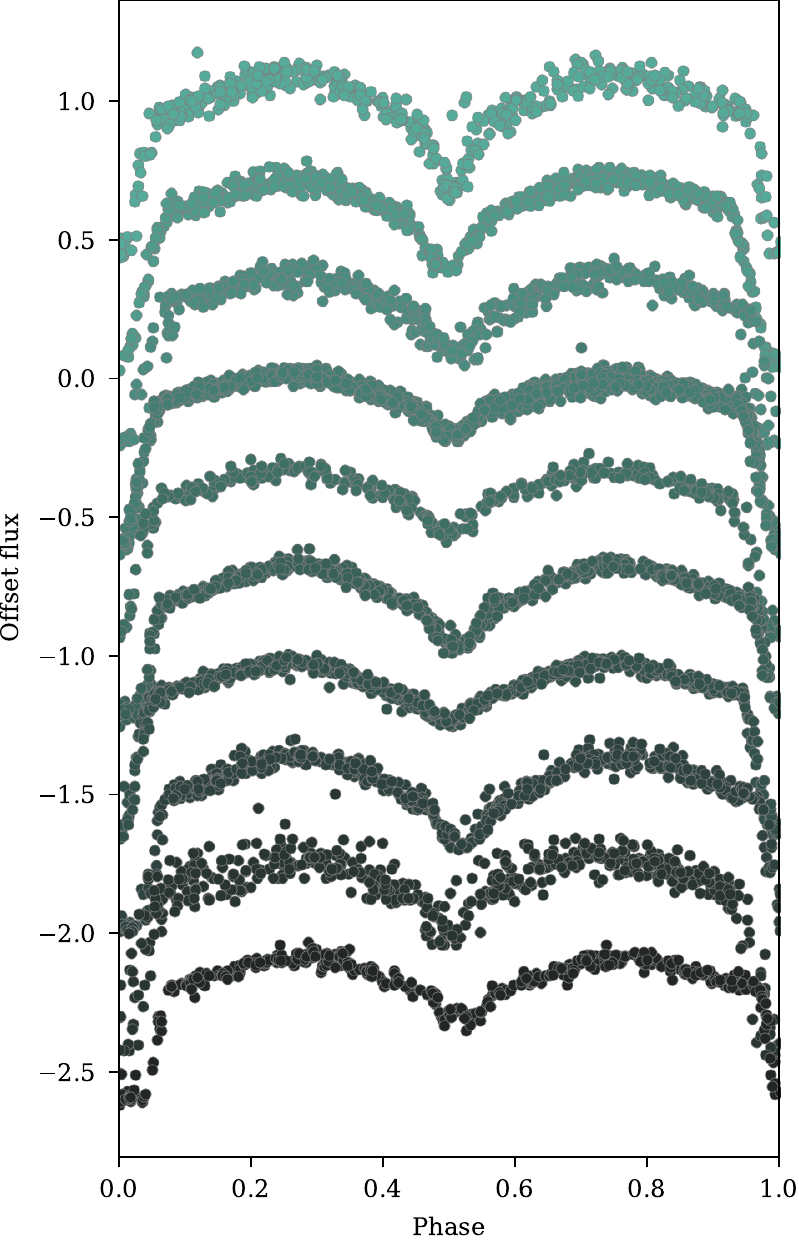}
\caption{A selection of several random eclipsing binaries with g-band photometry.}
\label{fig:ebs}
\end{figure}

To demonstrate the photometric precision of the ASAS-SN $g$-band photometry, we downloaded 1,000,000 random light curves from the stellar\_main catalogue, distributed over a range of Gaia DR2 magnitudes. For each light curve, we removed bad fluxes as indicated by the quality flags. We additionally normalized the flux of each camera to unity and perform a 5$\sigma$ clipping. We then measured the standard deviation of the entire light curve for each star. This metric is useful as a rough indication of the photometric precision for the source. We show the photometric precision of the $g$-band photometry as a function of Gaia magnitude in Fig.~\ref{fig:sat_range}.  

As seen in Fig.~\ref{fig:sat_range}, saturation in $g$-band begins to set in around $g \sim 12.5$ mag with the photometric precision becoming significantly worse around $g \sim 11$ mag.  At the time of this publication, we suggest using Sky Patrol V1.0 for significantly saturated ($g \lesssim 11.5$ mag) targets to avoid the residuals that image subtraction creates for such sources.  Future data releases we attempt to improve photometry for extremely saturated sources ($5 \lesssim g \lesssim 11.5$ mag).

We next illustrate the utility of ASAS-SN photometry with several non-transient variable sources. For each source, we use only the data returned by the query, with simple cuts on the quality flags as described above. Fig.~\ref{fig:saturated_ceph} shows the photometry of SV Vul, a bright ($g \approx 8.2$ mag; \citealp{tonry18}) classical Cepheid variable. While there are clear outliers that are not accurately captured by the data quality flags, there is no issue in recovering the known periodicity of around 45 days despite its brightness. 

Figure.~\ref{fig:RRLyr} demonstrates the exceptional baseline provided by ASAS-SN for an RR Lyrae variable star. Many RR Lyrae are known to undergo a quasi-periodic modulation of their variability over time, referred to as the Tseraskaya-Blazhko effect (\citealp{Buchler_2011}). Even in just the ASAS-SN $g$-band since 2017, this effect is easily visible. 

Finally, Fig.~\ref{fig:ebs} presents a sample of 10 randomly selected eclipsing binaries found by \citet{rowan22}.

% Unlike the previous iteration of Sky Patrol, it is now straight forward to perform science on an ensemble of stars from any combination of available catalogs in the database. 

\section{Database Metrics}
\begin{figure*}[t!]
\centering
\includegraphics[width=6.5in]{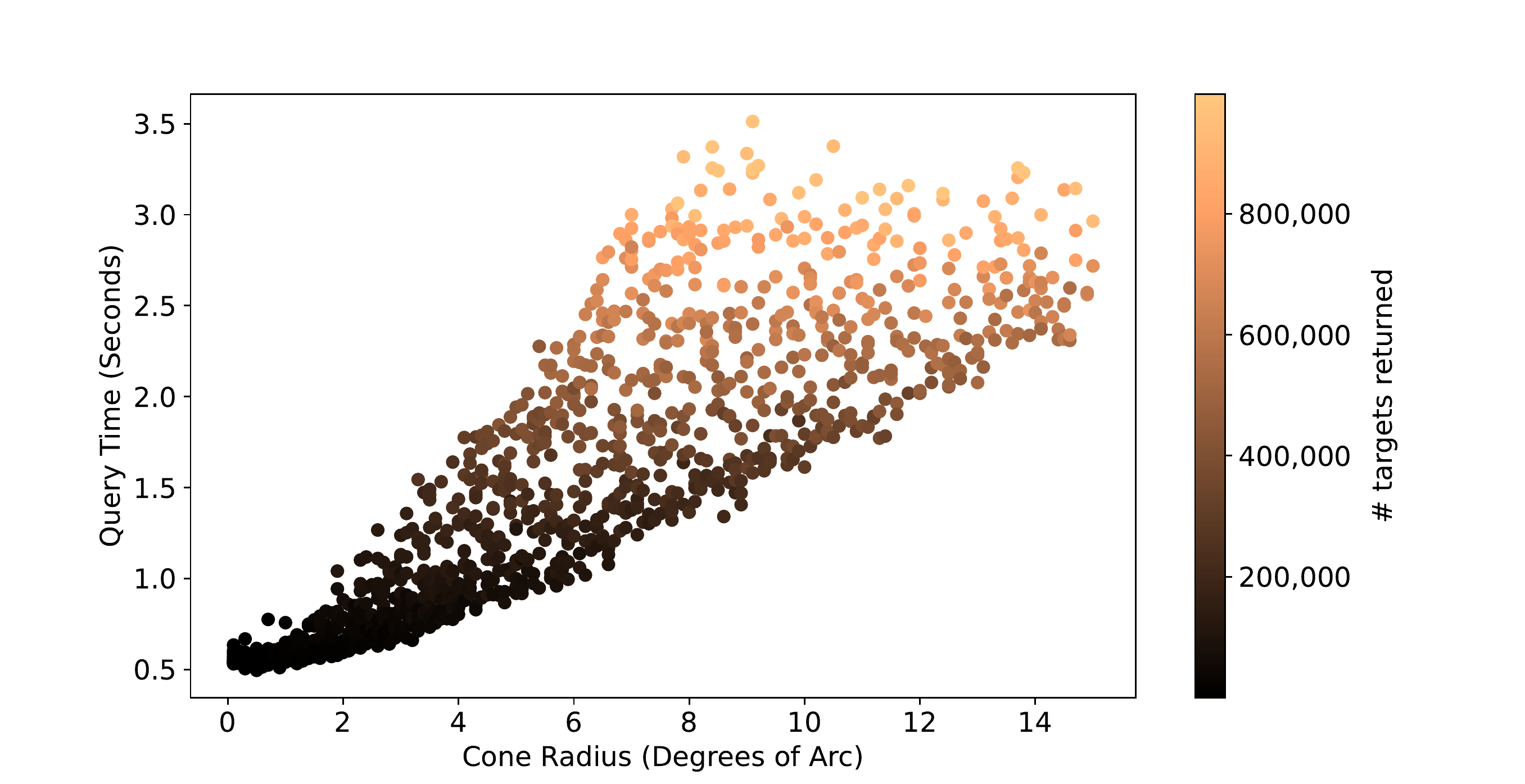}
\caption{Cone search lookup speed as a function of radius. Cone centers are sampled at random points across the sky.
  }
\label{fig:metrics}
\end{figure*}
The construction of this database presented us with a unique set of problems. Traditional survey data releases have been of static data sets that lend themselves to indexing and partitioning schemes that allow for fast queries. However, Sky Patrol V2.0 is not a static data release. Our light curves are updated in near-real time as we gather new images from our stations. This has forced us to decouple our lookup tables from their corresponding light curves. The lookup tables are kept as in-memory distributed dataframes and the light curves are kept on disk in a document store. Each of these architectures has its own contribution to the speed of our database. 

Our in-memory lookup tables do not have pre-calculated indexes on any of their columns. This is a restriction of the software stack we are running. However, this means that we can load new catalogs on-the-fly with little penalty. Also, it means that each one of our queries results in a full table scan. While this would be untenable on disk-based storage, it is trivial for in-memory storage. The major benefit of this is that the speed of our queries is not dictated by the complexity or breadth of their filters, but rather by the number of results they return (i.e., bandwidth limitations). In other words, users can run cone searches with arbitrary radius with little to no penalty. 

A major design goal of this architecture was to return all large queries to the lookup table---around 1 million targets---in under a minute, and all small queries---thousands of targets---in under 5 seconds. Figure 3 shows the cone search lookup speeds as a function of radii.

The document store uses our unique internal identifiers as hash keys for each corresponding light curve. This means we can retrieve light curves in $\mathcal{O}(1)$ time, so that retrieval time grows roughly with the number of sources. The speed of these retrievals is only limited by bandwidth.  Because the light curve documents contain far more data than their corresponding rows in the lookup tables, retrieval speeds will vary depending on the users bandwidth and latency from our servers at the University of Hawaii. In testing, downloads to Ohio State University ran at a rate of 1,000 light curves per minute per core.
\newpage
\section{Conclusion}

ASAS-SN Sky Patrol V2.0 represents a major step forward for survey data releases. While previous tools and data releases have given users the ability to pull live photometry data or run complex searches across a collection of input catalogs, few, if any, have managed to do both at the speed and scale of this project. 
While Sky Patrol V1.0 still allows users to run small numbers of light curves anywhere on the sky, Sky Patrol V2.0 enables users to perform studies using millions of light curves that are continuously updated.

It is our hope that researchers can leverage our input catalogs and light curves not only in their multi-messenger and time-domain applications, but also in service of standalone science applications. In regards to the latter, this system will serve as the foundation for the ASAS-SN team's upcoming ``Patrol'' projects. Each of these patrols will monitor incoming photometry data for certain classes of objects---such as active galactic nuclei, young stellar objects, cataclysmic variables, etc.---in order to detect and study anomalous events in real time.  Finally, we are also investigating the possibility of allowing individual users to create custom patrols where the target list and triggering function would be created by community members and run on the ASAS-SN data stream in real time. 

\acknowledgements{ACKNOWLEDGEMENTS}

We thank Las Cumbres Observatory and its staff for their continued support of ASAS-SN. ASAS-SN is funded in part by the Gordon and Betty Moore Foundation through grants GBMF5490 and GBMF10501 to the Ohio State University, and also funded in part by the Alfred P. Sloan Foundation grant G-2021-14192. Development of ASAS-SN has been supported by NSF grant AST-0908816, the Mt. Cuba Astronomical Foundation, the Center for Cosmology and AstroParticle Physics at the Ohio State University, the Chinese Academy of Sciences South America Center for Astronomy (CAS- SACA), and the Villum Foundation. 

We thank Roberto Assef, David Bersier, Laura Chomiuk, Xinyu Dai, Anna Franckowiak, JJ Hermes, Ondrej Pejcha, Sarah J. Schmidt, Jay Strader, and Maximilian Stritzinger for discussions during external catalog construction.  We also thank Chris Ashall, Thomas de Jaeger, Aaron Do, Jason Hinkle, and Anna Payne for other useful discussions during development.

KH and BJS are supported by NASA grant 80NSSC19K1717.  KH was also supported by the University of Hawaii Data Science Institute. BJS, CSK, and KZS are supported by NSF grant AST-1907570. BJS is also supported by NSF grants AST-1920392 and AST-1911074. CSK and KZS are supported by NSF grant AST-1814440. KZS is also supported by the 2022 Guggenheim Fellowship, and his stay at the UCSB KITP was supported in part by the National Science Foundation under Grant No. NSF PHY-1748958. Support for TJ was provided by NASA through the NASA Hubble Fellowship grant HF2-51509 awarded by the Space Telescope Science Institute, which is operated by the Association of Universities for Research in Astronomy, Inc., for NASA, under contract NAS5-26555.
KAA is supported by the Danish National Research Foundation (DNRF132).  MAT acknowledges support from the DOE CSGF through grant DE-SC0019323.  JFB was supported by NSF grant No. PHY-2012955. Support for JLP is provided in part by FONDECYT through the grant 1151445 and by the Ministry of Economy, Development, and Tourism's Millennium Science Initiative through grant IC120009, awarded to The Millennium Institute of Astrophysics, MAS. TAT acknowledges support from Scialog Scholar grant 24215 from the Research Corporation, a Simons Foundation Fellowship, and an IBM Einstein Fellowship from the Institute for Advanced Study, Princeton, while a portion of this work was completed.
Parts of this research were supported by the Australian Research Council Centre of Excellence for All Sky Astrophysics in 3 Dimensions (ASTRO 3D), through project number CE170100013.

\newpage
\bibliographystyle{mnras}
\bibliography{references}

\end{document}